\documentstyle[aps,prl,epsf,multicol]{revtex}
\begin{document}
\draft

\title{Excited Eigenstates and Strength Functions for Isolated Systems of
Interacting Particles}

\author{V.V.Flambaum$^{1}$ and 
F.M.Izrailev$^{2}$
}

\address{$^1$ School of Physics, University of New South Wales,
Sydney 2052, Australia}

\address{$^2$
Instituto de F\'isica, Universidad Autonoma de Puebla, Apdo. Postal J-48,
Puebla, 72570 M\'exico
}

\date{\today}
\maketitle

\begin{abstract}

Eigenstates in finite systems such as nuclei, atoms, atomic clusters and
quantum dots with few excited particles are chaotic superpositions of shell
model basis states. We study criterion for the equilibrium distribution of
basis components (ergodicity, or Quantum Chaos), effects of level density
variation and transition from the Breit-Wigner to the Gaussian shape of
eigenstates and strength functions. In the model of $n$ interacting
particles distributed over $m$ orbitals, the shape is given by the
Breit-Wigner function with the width in the form of gaussian dependence on
energy.


\end{abstract}

\pacs{PACS numbers:  05.45.+b, 31.25.-v, 31.50.+w, 32.30.-r}
\begin{multicols}{2}


{\bf Introduction.} Recently, based on chaotic structure of eigenstates,
statistical approach has been developed in \cite{F,FGGK94,FGI96,FI97} which
allows to find distribution of occupation numbers for single-particles
states, expectation values of different operators, transition amplitudes
between chaotic states and degree of an enhancement of weak interactions.
The main quantity in this approach is the form of the distribution of shell
model basis components in chaotic eigenstates. In recent numerical studies
of the Ce atom \cite{FGGK94}, the $s-d$ nuclear shell model \cite{ZELE} and
random two-body interaction model \cite{FGI96,FIC96} it was found that
typical shape of exact eigenstates practically does not depend on a
particular many-body system and has a universal form which essentially
depends on few parameters.

In this paper, we develop a method for the description of the form of
chaotic eigenstates and strength functions without diagonalization of huge
many-body Hamiltonian matrices.

We follow the standard way according to which the total Hamiltonian for $n$
Fermi-particles distributed over $m$ single-particles states is written as a
sum of two terms, 
\begin{equation}
\label{H}H=H_0+V=\sum_{s=1}^m\epsilon _sa_s^{+}a_s+\frac 12\sum
V_{pqrs}a_p^{+}a_q^{+}a_ra_s
\end{equation}
Here the ``unperturbed'' Hamiltonian $H_0$ incorporates an effect of the
mean field, $\epsilon _s$ are the energies of single-particle states
calculated in this field, $a_s^{+},a_s$ are creation-annihilation operators,
and $V$ stands for the residual two-body interaction (the difference between
the exact and mean field Hamiltonians).

Exact (``compound'') eigenstates $\left| i\right\rangle \,$of the
Hamiltonian $H$ can be expressed in terms of simple shell-model basis states 
$\left| k\right\rangle \,$(eigenstates of $H_0$) :

\begin{equation}
\label{slat}\left| i\right\rangle =\sum\limits_kC_k^{(i)}\left|
k\right\rangle \,;\,\,\,\,\,\,\,\,\left| k\right\rangle
=a_{k_1}^{+}...a_{k_n}^{+}\left| 0\right\rangle 
\end{equation}
These compound eigenstates $\left| i\right\rangle $ are formed by the
residual interaction $V$ ; in complex systems they typically contain large
number $N_{pc}\gg 1$ of the so-called principal components $C_k^{(i)}$ which
fluctuate ``randomly'' as a function of indices $i$ and $k$. The shape of
exact eigenstates is given by the ``spreading function'' $F$ (in what
follows, the $F-{\em function}$ ), 
\begin{equation}
\label{Fk}F_k^{(i)}\equiv \overline{\left| C_k^{(i)}\right| ^2}  \simeq
F\,(E_k, E^{(i)}) 
\end{equation}
where $E_k$ is the unperturbed energy and $E^{(i)}$ is the perturbed one.

{\bf Equations for strength function and spreading width.} For a weak
interaction between particles the shape of chaotic eigenstates is known to
be well described by the Breit-Wigner form. However, with an increase of the
interaction strength, the average shape of the eigenstates ($F-$function)
changes from the Breit-Wigner one to that close to the Gaussian with the
exponential tails \cite{FGGK94,FI97}. In order to reduce the distortion
effect due to non-constant density of states, in what follows we consider
the so-called ``strength function'' $P_k(E)$ which is also known in
literature as ``local spectral density of states'' (LDOS), 
\begin{equation}
\label{Ff}P_k(E)\equiv F(E_k,E)\rho (E)
\end{equation}
Note that the $F-$function  gives the shapes of both exact eigenstates and
the LDOS, depending on what is fixed, the total energy $E^{(i)}$ or the
unperturbed one, $E_k$.

The equations for $P_k(E)$ can be obtained on the base of the method
presented in \cite{BM69,Lau95}. Let us choose some basis component $\left|
k\right\rangle \,$ and diagonalize the Hamiltonian matrix without this
component. Thus, the problem is reduced to the interaction of this component
with the exact eigenstates $\left| i\right\rangle \,$ described by the
matrix elements $V_{ki}$.

The solution can be written by making use of the average over a small energy
interval $\Delta$ (see details in \cite{BM69}), 
\begin{equation}
\label{FfBW} P_k(E) = \frac{1}{2\pi}\frac{\Gamma_k(E)}{(E_k+\delta_k-E)^2 +
(\Gamma_k(E)/2)^2} 
\end{equation}
\begin{equation}
\label{Gamma} \Gamma_k(E) \simeq 2\pi \overline{\left| V_{ki}\right| ^2} 
\rho(E) 
\end{equation}
\begin{equation}
\label{delta}  \delta_k=\sum_i \frac{\left| V_{ki}\right| ^2 (E-E^{(i)})}
{(E-E^{(i)})^2 + (\Delta/2)^2} 
\end{equation}

It is easy to recognize in the energy shift $\delta _k$ the modified second
order correction to the unperturbed energy level. For the calculation of the
shape of the eigenvector $|i>$ one should substitute the exact energy $%
E=E^{(i)}=E_i+\delta _i$. The difference $\delta _i-\delta _k$ is negligible
if the interaction is not very strong.

One should stress that the summation in the above equations are performed
over exact states. We would like to express the result in terms of the basis
states only, this allows us to solve the problem without diagonalization of
the  Hamiltonian matrix $H_{kp}$. To do this, we express exact eigenstates $%
|i>$ in terms of the basis components, 
\begin{equation}
\label{Vki} |V_{ki}|^2 = \sum\limits_p\left| C_p^{(i)}\right| ^2 |H_{kp}|^2
+\sum\limits_{p\neq q} C_q^{(i)*} C_p^{(i)} H_{kp} H_{qk} 
\end{equation}
with $H_{kp}$ standing for non-diagonal Hamiltonian matrix elements. If
coefficients $C_{p,q}^{(i)}$ can be treated as random variables, the second
term vanishes after averaging. Substitution of Eq. (\ref{Vki}) into Eqs.(\ref
{Gamma}, \ref{delta}) gives 
$$
\Gamma_k(E)=2\pi \sum\limits_{p\neq k} |H_{kp}|^2 P_p(E)=%
$$
\begin{equation}
\label{Gammaf}  \sum\limits_{p\neq k} |H_{kp}|^2 \frac{\Gamma_p(E)}{%
(E_p+\delta_p-E)^2 + (\Gamma_p(E)/2)^2} 
\end{equation}
$$
\delta_k= \sum\limits_{p\neq k} \left| H_{kp}\right|^2 \int dE^{(i)} \frac
{P_p(E^{(i)})} {E-E^({i)}} 
\simeq%
$$
\begin{equation}
\label{deltaf} \sum\limits_{p\neq k} \frac{\left| H_{kp}\right| ^2 (E-E_p -
\delta_p)} {(E-E_p - \delta_p)^2 + (\Gamma_p(E)/2)^2} 
\end{equation}
where the integral is taken as the principal value. Last equality is valid
in the approximation of slow variation of $\Gamma_p(E)$ and $\delta_p$. The
equations for $\Gamma_k(E)$ and $\delta_k$ allow to calculate the strength
function (\ref{FfBW}) from the unperturbed energy spectrum and matrix
elements of the total Hamiltonian $H$.

{\bf Criterion of statistical equilibrium.} Let us discuss the conditions
under which self-consistent solution of the equations (\ref{FfBW}, \ref
{Gammaf}, \ref{deltaf}) exists. There are four important parameters in this
problem: the spreading width of a basis component $\Gamma $, the effective
band width $\sigma $ of the Hamiltonian matrix $H_{pq}$, the interval
between the many - body energy levels $D=\rho ^{-1}$ and the interval $%
d_f=\rho _f^{-1}$ between the final basis states $|p>$ which can be
connected with a particular basis component $|k>$ by the two-body
interaction. The ratio $D/d_f$ is exponentially small \cite{FGI96} since all
the basis states $|p>$ which differ from $|k>$ by position of more than two
particles, have zero matrix elements $H_{kp}$ and do not contribute to the
Eqs. (\ref{Gammaf}, \ref{deltaf}).

First, let us consider Eqs.(\ref{Gammaf}, \ref{deltaf}) for a strong enough
interaction, $\Gamma >>d_f$. In this case the number of effectively large
terms in the sums is large, $N_f\sim \Gamma /d_f$, and fluctuations of $%
\Gamma $ are small, $\delta \Gamma \sim \Gamma /\sqrt{N_f}$. Therefore, Eq. (%
\ref{Gammaf}) can be written as 
\begin{equation}
\label{GammaH}\Gamma _k(E)\simeq 2\pi \overline{\left| H_{kp}\right| ^2}\rho
_f(\tilde E)
\end{equation}
where $\tilde E=E-\delta $. The energy shift $\delta \equiv <\delta _p>$ can
be neglected in the case of $\Gamma <<\sigma $. In order to perform the
summation over $p$, we assumed that $\Gamma (E)$ and $\rho _f(E)$ vary
slowly within the energy interval of the size $\Gamma $. Thus, in order to
have large number of final states $N_f\sim 2\pi H_{kp}^2/d_f^2$ and
statistical equilibrium (small fluctuations of $\Gamma $), one needs $%
H_{kp}>>d_f$. In this case chaotic components of exact eigenfunctions in the
unperturbed many-particle basis ergodically fill the whole energy shell of
the width $\Gamma $, with Gaussian fluctuations of the coefficients $%
C_k^{(i)}$ with the variance given by the $F-$function (\ref{Fk}) (see also 
\cite{BM69,FGGK94}).

With the decrease of the ratio $H_{kp}/d_f$ the fluctuations of $\Gamma $
increase and for $H_{kp}<d_f$ the smooth self-consistent solution of Eqs.(%
\ref{Gammaf}) disappears. Indeed, in this case $\Gamma _p$ in the
denominator can be neglected and the sum in (\ref{Gammaf}) is dominated by
one term with a minimal energy $E-E_p\sim d_f$. Therefore, for a typical
basis state $|k>$ formally one gets $\Gamma _k\sim \Gamma
_p(H_{kp}/d_f)^2<<\Gamma _p$. This contradicts to the equilibrium condition
according to which all components are ``equal'' ( $\Gamma _k\sim \Gamma _p$).

One should stress that the absence of a smooth solution for the shape of the
eigenstates and the strength function does not mean that the number of
principal components in exact eigenstates is small. However, the
distribution of the components is not ergodic: there are many ``holes''
inside exact eigenstates which occupy the energy shell of the width $2\pi 
\overline{\left| H_{kp}\right| ^2}\rho _f(E)$ (see \cite{AGKL97,FI97}). In
such a situation, very large (non-Gaussian) fluctuations of $C_k^{(i)}$ are
typical.

It is important that ensemble averaging in this problem is not equivalent to
the energy average for a specific Hamiltonian matrix. For example, the
average over the single-particle spectrum leads to variation of energy
denominators in (\ref{Gammaf}) and can fill the holes in the $F-$function
even for $\Gamma < d_f$.

{\bf Transition from the Breit-Wigner type to the Gaussian-like strength
function.} In principal, the set of equations (\ref{Gammaf}, \ref{deltaf})
for the shape of the strength function $P_k(E)$ defined by Eq.(\ref{FfBW}),
can be solved numerically having the unperturbed many-body spectrum and
matrix elements $H_{kp}$ of the total Hamiltonian. However, for relatively
large number of particles (practically, for $n\ge 4$), one can find an
approximate analytical solution of the problem.

First, we note that the spreading width $\Gamma (E)$ in the expression (\ref
{FfBW}) for the strength function can be a strong function of excitation
energy $E$ due to the variation of the density of the final states $\rho
_f(E))=(d_f)^{-1}$ in Eq. (\ref{GammaH}). It is well-known that at small $E$
the basis component with one excited particle has $\Gamma (E)\propto E^2/d_0$
where $d_0$ is the interval between single-particle energy levels. For
typical case of $n^{\star }\sim (E/d_0)^{1/2}$ excited particles the
spreading width can be estimated as $\Gamma (E)\propto (d_f)^{-1}\sim
(d_0)^{-1}(E/d_0)^{3/2}$ \cite{FGS97}. Below we show that at higher energies
far from the ground state, the energy dependence of $\rho _f(E)$ and $\Gamma
(E)$ can be quite close to the Gaussian. Note that the Gaussian form
typically occurs in ``statistical spectroscopy'' \cite{FW70} when neglecting
the mean field term in Eq.(\ref{H}).

In the model (\ref{H}) the density $d_f$ is defined by transitions between
those basis states which differ by the position of one or two particles
only, therefore, $\rho_f(E) = \rho_f^{(1)}(E) + \rho_f^{(2)}(E) $.

Let us estimate the density $\rho_f^{(2)}$ determined by the energy
difference $\omega_{pk}^{(2)}$ between the states $|p>$ and $|k>$ which
differ by the position of two particles, 
$$
\omega_{pk}^{(2)}=\epsilon_{\alpha}^{(p)}+ \epsilon_{\beta}^{(p)} -
\epsilon_{\gamma}^{(k)} -\epsilon_{\delta}^{(k)} 
$$
\begin{equation}
\label{omegaPK} +\sum\limits_{\nu \neq \alpha , \beta , \gamma ,\delta} 
(V_{\alpha \nu} + V_{\beta \nu} -V_{\gamma \nu} - V_{\delta \nu}) +
V_{\alpha \beta} - V_{\gamma \delta} 
\end{equation}
Here the summation is taken over $n-2$ occupied orbitals and $V_{\alpha \nu}$
are the diagonal matrix elements of the residual interaction between the
particles located at the orbitals $\alpha$ and $\nu$. The matrix elements of
residual interaction are assumed to be random with the zero mean.

For large number of fluctuating terms in the Eq.(\ref{omegaPK}) the
distribution of $\omega $ is close to the gaussian form. Strictly speaking,
this is correct if the contribution of $4n-6$ interaction terms to the
frequency Eq.(\ref{omegaPK}) is strong. However, even four single-particle
energy terms give the distribution which is close to the Gaussian. The same
conclusion is reasonable also for single-particle transition density $\rho
_f^{(1)}$, thus, the general expression reads as 
\begin{equation}
\label{rhof}\rho _f^{(1,2)}(\tilde E)\simeq K(2\pi \sigma ^2)^{-1/2}exp(-
\frac{(\tilde E-E_k-\overline{\omega })^2}{2\sigma ^2}).
\end{equation}
The normalization parameter $K$ stands for the number of one or two-particle
transitions, $K=K_1=n(m-n)$ and $K=K_2=n(n-1)(m-n)(m-n-1)/4$ correspondingly 
\cite{FGI96}.

From Eq.(\ref{omegaPK}) the estimate for the average frequency of
two-particle transitions reads as $\overline{\omega ^{(2)}}=2(\overline{%
\epsilon _p}-\overline{\epsilon _k})\approx 2m/(m-n)(\overline{\epsilon }%
-E_k/n)$ where $\overline{\epsilon _k}=E_k/n$ is the average single-particle
energy in the basis state $|k>$ containing $n$ particles, $\overline{%
\epsilon }$ is the single-particle energy averaged over all $m$ orbitals.
Average energy of the empty orbitals $\overline{\epsilon _p}$ can be found
from the relation $m \overline{\epsilon }=\overline{\epsilon _k}n+\overline{%
\epsilon _p}(m-n)$.

The variance of $\rho _f^{(2)}(E)$ for two-particle transitions is equal to $%
\sigma _2^2=2\sigma _p^2+2\sigma _k^2+(4n-6)V^2\approx 2\sigma _\epsilon
^2+(4n-6)V^2$ where $\sigma _\epsilon ^2$ is the variance of single-particle
spectrum, and $V^2$ is the variance of non-diagonal matrix elements of the
two-body residual interaction. Note that in the case of $n<<m$ for low-lying
states the variance of the occupied orbital energies $\sigma _k^2$ is small
and the variance of empty orbital energies is $\sigma _p^2\sim \sigma
_\epsilon ^2$.

Similar, the density $\rho_f^{(1)}$ is also approximated by the Eq.(\ref
{rhof}), with $K=K_1$ , $\overline{\omega^{(1)}} \approx m/(m-n)( \overline{%
\epsilon} - E_k/n) $ and $\sigma_1^2= \sigma_p^2 + \sigma_k^2 + 2(n-1)V^2
\approx \sigma_{\epsilon}^2 + 2(n-1)V^2 $.

Thus, the width $\Gamma (E)$ is given by the following expression, $\Gamma
=2\pi [(n-1)V^2\rho _f^{(1)}+V^2\rho _f^{(2)}]$. Since for single-particle
transitions the summation in $H_{kp}=\sum_\nu V_{\alpha \nu \rightarrow
\gamma \nu }$ is performed over occupied orbitals, the factor $n-1$ appears
in the above relation. Typically, the ratio $K_2/[(n-1)K_1]=(m-n-1)/4$ is
larger than 1, therefore, the two-particle transitions dominate. In this
case we can neglect the differences in $\overline{\omega }$ and $\sigma $
for two-particle and one-particle transitions and combine two terms into
one. As a result, the spreading width is described by the simple Gaussian
formula 
\begin{equation}
\label{GammaG}\Gamma _k(E)\simeq 2\pi (\Delta E)_k^2\frac 1{\sqrt{2\pi
\sigma _k^2}}exp\left\{ -\frac{(\tilde E-E_k-\overline{\omega _k})^2}{%
2\sigma _k^2}\right\} 
\end{equation}
where $\tilde E=E-\delta $. Here $(\Delta E)_k^2$ is the variance of the
strength function which can be defined through its average value \cite
{FGI96,FI97}, 
$$
\overline{(\Delta E)_k^2}=\overline{\sum_{p\neq k}H_{kp}^2}%
=V^2n(n-1)(m-n)(3+m-n)/4 
$$
and $\omega _k$ and $\sigma _k$ are close to that for the two-particle
transitions. The maximum of $\rho _f(E)$ and $\Gamma (E)$ is shifted by $|
\overline{\omega _k}|$ towards the center of the spectrum compared to the
maximum of Breit-Wigner function. This leads to some distortion of the
strength function Eq.(\ref{FfBW}) and the shape of the eigenstates, which is
especially large at the bottom of the spectrum.

Thus, we have demonstrated that if the interaction is small ($\Gamma \ll
\sigma _k$), the strength function has the Breit-Wigner shape with the broad
gaussian envelope originating from $\Gamma _k(E)$ in the numerator of Eq. (%
\ref{FfBW}). It is easy to check that this envelope is, indeed, needed in
order to provide the correct value $(\Delta E)_k^2$ for the second moment of
the strength function (note, that the Breit-Wigner shape has infinite second
moment which is unphysical).

When the interaction $V$ increases one needs to take into account one more 
contribution to the broadening of the shape of $\Gamma(E)$. It is given by
the width of the strength function $P_p(E)$ in Eq.(\ref{Gammaf}) (it was
neglected in Eq.(\ref{GammaH})). Taking into account this width we can give
an estimate $\sigma_k^2 \simeq \sigma_2^2 + \overline{\Gamma_p^2}$. With
further increase of interaction, where the shape of $P_p(E)$ is close to the
Gaussian, we have $\sigma_k^2 \simeq \sigma_2^2 + (\Delta E)_k^2$.

Direct numerical study of the model (\ref{H}) with $n=6$ Fermi-particles and 
$m=13$ orbitals shows that the above analytical expressions give quite a
good description of the shape of the strength function $P_k(E)$ as well as
of the energy dependence of the spreading width. The unperturbed
single-particle spectrum has been chosen at random, with $d_0=1$ and $%
\epsilon _s\approx d_0s$. The size of the Hamiltonian matrix is $N=C_m^n=1716
$ and we specify the unperturbed state $|i_0>$ with $i_0=280$.

\begin{figure}
\vspace{0.1cm}
\hspace {-1.1cm}
\epsfxsize 8cm
\epsfbox{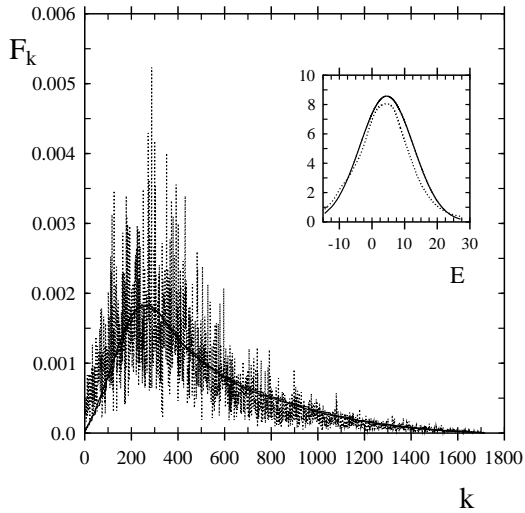}
\vspace{-0.4cm}
\narrowtext
\caption{
The $F-$function (\ref{Fk}) in the basis representation. Broken line is the
result of numerical diagonalization of the Hamiltonian matrix; to reduce the
fluctuations, we take the average over 10 matrices $H_{ik}$ with different
two-body random interaction with $V^2\approx 0.15$. Two smooth curves
correspond to the computation of Eq.(\ref{Ff}) with $\Gamma _k(E)$ given by
Eqs.(\ref{Gammaf},\ref{deltaf}) and by Eq.(\ref{GammaG}); they practically
coincide. The inset shows the dependence $\Gamma _k(E)$ itself; full curve
is the expression (\ref{GammaG}), the dashed curve is the computation from
Eqs.(\ref{Gammaf}, \ref{deltaf}).
}
\label{g}
\end{figure}

In the estimates above we assumed that $\Gamma <<\sigma $ where $\sigma $ is
the effective energy band width of the Hamiltonian matrix, see Eq.(\ref
{GammaG}). When $\Gamma \sim \sigma $, the (Gaussian) variation of $\Gamma
(E)$ in the numerator of the strength function in Eq.(\ref{FfBW}) becomes as
important as the variation of the Breit- Wigner energy denominator $%
(E-E_k)^2+(\Gamma /2)^2$. At this point, $\Gamma \approx \sigma $, the
transition from the Breit-Wigner type to Gaussian-type shape of the
eigenstates takes place. We still can use Eqs.(\ref{Gammaf}, \ref{deltaf}),
and (\ref{Ff}) in order to calculate (numerically) $\Gamma (E),P_k(E)$ and $%
F(E,E_k)$, using $\Gamma $ from Eq.(\ref{GammaG}) with $\sigma _k^2\simeq
\sigma _2^2+(\Delta E)_k^2$ as the zero approximation in the right-hand side
of Eqs.(\ref{Gammaf}, \ref{deltaf}).

{\bf Acknowledgements.} VVF is grateful to ICTP (Triest), Princeton
University and NEC Research Institute for hospitality. VVF acknowledges the
support from Australian Research Council, and FMI acknowledges the support
from the CONACyT Grant No. 26163-E (Mexico).
Both authors are very thankful to Prof. V.Zelevinsky
for valuable comments and kind hospitality during the stay in MSU Cyclotron
laboratory when this work was completed.

\end{multicols}
\end{document}